\documentclass[apj,twocolumn]{emulateapj}
\usepackage{amsmath}
\usepackage{psfig}
\usepackage{graphicx}
\usepackage{multirow}
\usepackage{times}
\usepackage{float}
\usepackage{natbib}
\begin{document}
\title{Emergent Spectra From Disks Surrounding Kerr Black Holes: Effect of Photon Trapping and Disk Self-Shadowing}
\author{Center for Astrophysics, Department of Astronomy,
University of Sciences and Technology of China,
Hefei 230026, China}

\altaffiltext{3}{Key Laboratory for Research in Galaxies and
Cosmology, Shanghai Astronomical Observatory, Chinese Academy of
Sciences, Shanghai 200030, China; cxw@shao.ac.cn}

 \begin{abstract}

Based on a {new} estimation of their thickness, the global properties
of relativistic slim accretion disks are investigated in this work.
The resulting emergent spectra are calculated using the relativistic
ray-tracing method, in which we neglect the self-irradiation of the accretion disk.
The angular dependence of the disk luminosity,
the effects of the heat advection and the disk thickness
on the estimation of the black hole spin are
discussed. Compare to the previous works, our improvements are that we use the self-consistent disk equations
and we consider the disk self-shadowing effect. We find that at the moderate accretion rate,
%with inclusion of the heat advection effect,
the radiation trapped in the outer region
of the accretion disks will escape in the inner region of the
accretion disk and contribute to the emergent spectra. At the high
accretion rate, for the large inclination and large black hole spin, both
the disk thickness and the heat advection have significant influence
on the emergent spectra. Consequently, these effects will influence the
measurement of the black hole spin based on the spectra fitting and influence the angular dependence of the luminosity.
For the disks around Kerr black holes with $a=0.98$, if the disk inclination is greater than $60^\circ$, and their
luminosity is beyond 0.2 Eddington luminosity,
the spectral model which is based on the relativistic standard accretion disk is no longer applicable for the spectra fitting.
{ We also confirm that the effect of the self-shadowing is significantly enhanced by the light-bending,
which implies that the non-relativistic treatment of the self-shadowing is inaccurate.
According to our results, the observed luminosity dependence of the measured spin suggests that the disk self-shadowing
significantly shapes the spectra of GRS 1915+105, which might lead to the underestimation of the black hole spin for the
high luminosity states.}
 \end{abstract}

\normalsize
 \section{Introduction}
The standard accretion disk model
\citep[SSD,][]{1973A&A....24..337S} and its relativistic
generalization \citep{1973blho.conf..343N} are widely used in
modeling the spectral energy distribution (SED) of both active galactic nuclei (AGNs) and the galactic black hole binaries. Recently, based
on the relativistic SSD, the spins of the black holes in several
galactic black hole binaries are measured by fittings of their SED
\citep{2005ApJS..157..335L,2006ApJ...636L.113S,2006ApJ...652..518M,2006MNRAS.373.1004M}.
One major advantage
of standard accretion disk model is its simplicity, i.e., it assumes
that the fluid undergoes the Keplerian motion and the energy generated
by the viscous heating is radiated locally. Besides, in this model, the
disk terminates at the Innermost Stable {Circular} Orbit (ISCO), and
the matter inside ISCO is cold.

However, when the accretion rate is relatively high ($\dot M>0.1
\dot M_{\rm edd}$), in the inner region of the accretion disks,
several effects will make these assumptions invalid. One such effect
is the heat advection. When the vertical diffusion time scale is
larger than the accretion time scale, substantial amount of heat is
trapped inside the accretion flow and not able to escape to the infinity
\citep{1978MNRAS.184...53B,1982ApJ...253..873B}. This will make the
local flux different from that from the SSD model and affect the
profile of the radial temperature. Also, the advected heat will provide an
additional pressure, which makes the disk rotation no
 longer Keplerian.
 Besides, because the heat trapped inside
the accretion flow, the region inside ISCO is not cold any more.

The self-consistent treatment of these effects needs to model the
energy and momentum balance in the disk, one widely used method
is based on the radiative MHD simulation. Although the numerical simulations
have the advantage of
treating the physical effects self--consistently, the huge amount of
computation makes it difficult to give the qualitative fitting of
the observational data. One complementary approach is to solve the energy and
momentum equations of the accretion flow in terms of the ordinary
differential equations (ODEs). Compared to the simulations, this method
can effectively take into account the dynamical effects, while
maintaining its conciseness. In this work, we use the ODE form disk
equations
\citep{1981AcA....31..283P,1988ApJ...332..646A,1993ApJ...412..254C,1997MNRAS.286..681P,1998ApJ...498..313G},
and explore the solutions at both the medium and high accretion rate
\citep[where the disk is call the ``slim
disk";][]{1988ApJ...332..646A}. We use the relativistic disk equations
in our work
\citep{1996ApJ...471..762A,1998MNRAS.297..739B,2003MNRAS.338.1013S}
to investigate the effects of the black hole spin.

The physical properties of the black hole and the accretion disks are deduced mainly
from their observed SED. It is well known that the structure and the SED of the slim disk
differ significantly from that of
SSD \citep{1999ApJ...516..420W,1999ApJ...522..839W}.
The disk SED is influenced by a handful of
effects, such as the Doppler effect, the relativistic beaming, the
gravitational redshift, the gravitational -bending, and the disk
self-shadowing. As the accretion rate increases, the disk becomes
geometrically thick, and the disk
self-shadowing \citep{2007MNRAS.378..841W} becomes significant. In this work, we
employ a ray-tracing approach
\citep[e.g.][]{1997PASJ...49..159F,1998NewA....3..647C} to explore
 the final spectra at the different viewing angles.

Due to their disk-like geometry, the radiation from the accretion disks is far from the isotropy,
and the emergent luminosity is thus angular dependent \citep[e.g.][]{2006MNRAS.372.1208H}.
For a slab of optically thick material without velocity, the disk luminosity simply varies as $L\sim \cos\theta$
in the flat spacetime.
In fact, because of its thickness and rotation, the angular dependence of the disk
luminosity is expected to be different from that of the simple slab, further analysis is thus needed.
In this paper, we calculate  the angular dependence of the disk luminosity under different inclinations
and the black hole spin.

% Most of the observational studies of accretion disks focusing on
% spin estimation are based on the fitting of their SED at X-ray. One
% widely used model is the XSPEC model KERRBB
% \citep{2005ApJS..157..335L}. To show how the effects of the heat
% advection and the self-shadowing of the disk affect the spin
% estimation, we also compare our calculated SED with these from
% KERRBB.

{
The spin of GRS 1915+105 is debated in the literature \citep{2006ApJ...652..518M,2006MNRAS.373.1004M}.
By fitting the disk blackbody component, \citet{2006ApJ...652..518M} found that the spin
is near extreme, while \citet{2006MNRAS.373.1004M} found the spin is moderate, with $a \sim 0.8$.
To show how the spin determination is influenced by the heat advection and disk self-shadowing,
we re-simulate the observation carried out by \citet{2006ApJ...652..518M} with our model and
confirm the luminosity dependence of the measured spin as found by \citet{2006ApJ...652..518M}.
}

In this work, the black hole mass in our canonical model is taken to be $10 M_{\odot}$.
However, due to the physical similarities between the accretion disks around
the stellar-mass black holes and those around the supermassive black holes, our results can be
easily generalized to the supermassive black
holes and predict the SED of AGNs under the simple scaling.

This paper organized as follows. In section 2, we describe briefly our
disk model and the ray--tracing method. In Section\ref{secst}-Section
\ref{secend}, we present our numerical results of the global disk
structure. In Section \ref{respec}, we present our results on the SED of
the disk, focusing on the differences between the SED of the
dynamical disk model and that of the SSD. The angular dependence of
the total luminosity is given in Section \ref{angulardep}. In Section
\ref{reimp}, implications for estimating the black hole spin are discussed.  In Section \ref{disc}, various
observational implications of the disk self-shadowing as well as the spin of GRS 1915+105 are discussed.
Finally, in Section \ref{recon} we summarize our conclusions.

\section{The Relativistic Slim Model}
\subsection{The Dynamical Disk Model}
{
The basic equations governing the disk structure used in our work are similar to those given in \citet{2003MNRAS.338.1013S}.
% but with some modifications. For tness we list the equations below.
In the Boyer--Lindquist coordinates, expanding around the equatorial plane up to $(z/r)^0$ terms, the metric takes the form \citep[][the geometric unit $G=C=M=1$ is taken]{1973blho.conf..343N} :
}
\begin{equation}
ds^{2}=-r^{2}\frac{\Delta}{A}dt^2+\frac{A}{r^2}(d\phi-\omega dt)^2+\frac{r^2}{\Delta}dr^2+dz^2\;,
\end{equation}
with
$$\Delta=r^2-2r+a^2\;,$$
$$A=r^4+r^2 a^2+2 r a^2\;,$$
$$\omega=\frac{2 a r}{A}\;,$$
where $a$ is the dimensionless spin of the black hole.

 The mass conservation takes form
\begin{equation}\label{conservation}
 \dot M=2 \pi \Delta^{1/2}\Sigma \gamma_{r} V \;,
\end{equation}
where $\dot M$ is the mass accretion rate, and $W$ and $\Sigma$ are
integrated pressure and density, defined as $W=\int p\ d z
$, $\Sigma=\int \rho\ d z $. Here, $p$ means pressure and $\rho$ means density, the subscript ``$0$" denotes the
corresponding value at the equatorial plane, and $H$ is the scale height of the
accretion disk. $\gamma_{r}$ is the radial Lorentz factor of the
fluid.

The angular momentum conservation reads
\begin{equation}\label{angular}
\frac{\dot M}{2\pi}(L-L_{\rm in})=\alpha \frac{A^{3/2}\Delta^{1/2} \gamma_{\phi}^3}{r^5} W \;,
\end{equation}
where $L$ is the specific angular momentum, $L_{\rm in}$ is the
specific angular momentum at the inner boundary of the accretion flow, and
$\gamma_{\phi}$ is the Lorentz factor in the $\phi$ direction as
defined in \citet{2000ApJ...534..734M}.
{As the viscosity prescription used here is conventional \citep{1992ApJ...394..261N}, at the inner part of the accretion disk,
our result may be inaccurate in principle. However, the causal viscosity is still somewhat putative and differs from the conventional one only close to
or inside the ISCO. Thus, the disk spectrum in our Kerr case is robust in the sense that the sonic point is very close
to the event horizon of the black hole, while the disk spectrum in the Schwarzschild case may be shifted a bit
due to the different treatment of the prescription of the viscosity.
}

The momentum equation and energy equation take the forms as follows:\\
Momentum equation:
\begin{equation}
\gamma_{r}^2 V \frac{ d V}{{d} r}=-\frac{1}{\Sigma}\frac{\rm d W}{{d} r}-\frac{\gamma_{\phi}^2 A}{r^4\Delta}
\frac{(\Omega-\Omega_{\rm k}^+)(\Omega-\Omega_{\rm k}^-)}{\Omega_{\rm k}^+\Omega_{\rm k}^-}\;,
\end{equation}
where
$$\Omega_{\rm k}^{\pm}=\pm\frac{1}{r^{\frac{3}{2}}\pm a}\;$$ is the
Keplerian angular velocity of the direct (``$+$ sign'') and retrograde (``-''
sign) circular orbit, respectively. And here
$$\gamma_{\phi}=\Big(1+\frac{r^2 L^2}{\gamma_{r}^2
A}\Big)^{1/2}\;$$ is Lorentz factor of $v_{\phi}$,
$$\Omega=\omega+\tilde\Omega=\omega+\frac{r^3 \Delta^{1/2} L}{A^{3/2} \gamma_{r}\gamma_{\phi}}\;.$$
Energy equation:
\begin{equation}Q^+_{\rm vis}=Q^-_{\rm adv}+F_{\rm rad}^- \;,
\end{equation}
here, $Q^+_{\rm vis}$ is the viscous heating rate per unit surface,
$F^-_{\rm rad}$ is the radiative cooling term, and $Q_{\rm adv}^-$
is the effective cooling induced by heat advection. They can be
specified as follows:
\begin{equation}
 Q^+_{\rm vis}=-\frac{1}{4 \pi r}\frac{\gamma_{\phi} A^{1/2}}{\Delta^{1/2} r} \dot M (L-L_{\rm in})\frac{{d} \Omega}{{d} r}\;,
\end{equation}
\begin{equation}
\begin{split}
Q^-_{\rm adv}=-\frac{\dot M}{4 \pi r}\frac{W}{\Sigma}\frac{1}{\Gamma_3-1}\Big( \frac{{d} \ln W}{{d} r}- \\ \Gamma_1
\frac{{d} \ln \Sigma}{{d} r}-(\Gamma_1+1)\frac{{d} \ln H}{{d} r}     \Big)\;,
\end{split}
\end{equation}
where $\Gamma_1$ and $\Gamma_3$ are defined in
\citet{1998bhad.conf.....K}, and
\begin{equation}\label{Frad}
 F^-_{\rm rad}=\frac{8 a c T^4}{3 \kappa \rho_0 H}\;.
\end{equation}
The Rosseland mean of total opacity can be expressed as
$$\kappa=0.4+0.64\times10^{23}\rho T^{-7/2}\;.$$

The total pressure can be expressed as the sum of the gas and
radiation pressure. For simplicity, we drop the numerical factors in
\citet{2003MNRAS.338.1013S}, and the equation of state
is given by:
\begin{equation}
W=\frac{1}{3}a T^4 H+\frac{2 k_{\rm B}}{m_{\rm p}}\Sigma T\;,
\end{equation}
where $T$ is temperature of the disk.
 Another equation is the vertical thickness equation, which
reads
\begin{equation}\label{thick}
 H=\sqrt{12}\frac{c_{\rm s}}{\Omega_{\rm k}}g_{\rm pa}^{-\frac{1}{2}}\;.
\end{equation}
This is a modification to the equations of \citet{2003MNRAS.338.1013S}, and its derivation will be presented in Section \ref{VerticalThick}.

The above equations are transformed into a set of first-order
differential equations with variables $v_{\rm r}$ and $W$, and then
integrated from the outer boundary at about $10^5 r_{\rm g}$.

The transonic solutions \citep{1980ApJ...240..271L} are obtained
using a shooting method, in which we adjust $L_{\rm in}$ recursively
until the smooth transonic solution is found.
During the solving process, despite that the factor $g_{\rm pa}$
takes form
\begin{equation}
 g_{\rm pa} '=\gamma_{\phi}^2\Big (\frac{(r^2+a^2)^2+2 \Delta a^2}{(r^2+a^2)^2-\Delta
 a^2}\Big) \;,
\end{equation}
a simplified version of factor $g_{\rm pa}$ is used when finding the
global transonic solution:
\begin{equation}
 g_{\rm pa}=\frac{(r^2+a^2)^2+2 \Delta a^2}{(r^2+a^2)^2-\Delta
 a^2}\;.
\end{equation}
However, the location of the photosphere \citep{1997MNRAS.286..681P} is still
determined  according to Equation (\ref{phsphere}).
When this simplification is taken, at the inner regions of the
accretion disks, the resulting disk photosphere
and disk thickness no longer agree with each other. But the difference is not significant.
We thus neglect this inconsistency and use the location of the disk photosphere in the ray-tracing calculation.

\subsection{Estimation of the Vertical Thickness}
\label{VerticalThick} The vertical thickness is a major uncertainty
in the accretion disk theory, and is claimed to be dependent on
various physical assumptions such as the vertical dissipation profiles
and the interplay between the relaxation time scale and the free-fall
time scale. However, \citet{2009A&A...502....7S} showed that the
vertical thickness of an accretion disk in the non-advective regime
can be approximated using this simple formula:
\begin{equation}\label{phsphere}
\kappa \sigma T_{\rm eff}^4\frac{1}{c}=\frac{2}{3} \Omega_{\rm k}^2 z g_{\rm pa}\;,
\end{equation}
where $\kappa$ is the total opacity, $\sigma$ is Stefan--Boltzmann
constant and $z$ is the photosphere location. $T_{\rm eff}$ is the
effective temperature defined as $T_{\rm eff}=( F_{\rm rad}^- /
\sigma )^{1/4}$ and $\Omega_{\rm k}=r^{-3/2}$. Equation (\ref{phsphere}) simply means that at the
disk photosphere, the vertical radiation pressure balances the
gravity of the central black hole. In this work, despite the fact
that Equation (\ref{phsphere}) has only been shown to be valid in
the non-advective regime, we also use it in the advective regime,
 since the physical argument is still valid. With this assumption, we can get the expression 
for the location of the photosphere:
\begin{equation}\label{z}
 z=\kappa \frac{3 \sigma T_{\rm eff}^4 }{2 \Omega_{\rm k}^2 g_{\rm pa}}\frac{1}{c}
=\frac{3}{2}\kappa \frac{F_{\rm rad}}{c}\Omega_{\rm k}^{-2} g_{\rm pa}^{-1}\;.
\end{equation}

Another constraint comes from the vertical radiation flux. According to the diffusion approximation,
assuming the dominance of the  radiation pressure, the radiation flux can be approximated as
\begin{equation}\label{rad}
 F_{\rm rad}=\frac{8 a c T^4}{3 \kappa \rho_0 H}=\frac{8 c W}{\kappa \Sigma H}\;.
\end{equation}
We thus make the physical requirement that the thickness $z$
obtained in Equation (\ref{z}) should be consistent with the thickness $H$ in
Equation (\ref{rad}). Combining these two equations, we obtain the expression
for modified disk thickness:
\begin{equation}\label{thick}
 H=\sqrt{12}\frac{c_{\rm s}}{\Omega_{\rm k}}g_{\rm pa}^{-\frac{1}{2}}\;.
\end{equation}
It should be noted that
Equation (\ref{thick}) is accurate when
the disk is radiation pressure dominated. At the outer regions of
the accretion disks, where the gas pressure dominates, Equation (16) deviates
from the previous one by a factor of $\sqrt{12}$ (the previous expression takes form $ H=c_{\rm s}/\Omega_{\rm k} g_{\rm pa}^{-1/2}
$). However, as \citet{2009A&A...502....7S} showed, the thickness in the outer region of the accretion disks is not well constrained,
and it is dependent on the dissipation profile in the vertical direction. As the contribution of radiation from the
gas-dominated region is relatively unimportant and the final emergent SED is
unaffected by our modifications significantly, we use Equation(\ref{thick}) throughout the paper.
{One concern about the estimation of thickness is that the effect of magnetic field may influence the disk thickness.
\citet{2006ApJ...640..901H} pointed out that the location of photosphere of the accretion disk is modified by the magnetic
field significantly. However, in our case, the shadowing effect happens at the inner region, where the radiation pressures
dominates. Thus the thickness and the self-shadowing might not be significantly affected by the magnetic effect.
The magnetic field might also change the inner boundary condition significantly \citep[e.g.][and references therein]{2010arXiv1003.0966P,2010ApJ...711..959N}.
To investigate the effects of the magnetic field, we should do the general relativistic MHD simulations
\citep{2003ApJ...589..444G,2003ApJ...599.1238D}, which is beyond the
scope of this work.
}
\subsection{Ray-Tracing Method}
After obtaining the global structure of the relativistic disk, we
use the ray-tracing technique to calculate the emergent spectra with
inclusion of all the relativistic effects. The ray-tracing method is
based on the analytical integration of the null-geodesics
\citep{1983mtbh.book.....C,1994ApJ...421...46R,1998NewA....3..647C},
and the subroutines are the same as those used in
\citet{2009ApJ...699..722Y}.

One major improvement compared to earlier work \citep[e.g.
][]{2005ApJS..157..335L} is that we include the disk thickness in
the ray-tracing calculation. To achieve this, we start from the infinity, and
then search along the each photon trajectory for the intersection point.

Once the intersection points are found, we can calculate the local
temperature of the emitting points in our disk model and
calculate the redshift factor $g$. The emergent spectra are then
obtained by integrating over the photometric plane according to
\begin{equation}
 F_{\nu_o}=\int g^{3} I_{\nu_{e}} \frac{{d} \alpha {d} \beta}{D^2}\;,
\end{equation}
where $\alpha$ and $\beta$ are two impact parameters, and the redshift factor is
\begin{equation}
 g=\frac{\nu_{\rm o}}{\nu_{\rm e}}\;,
\end{equation}
here $D$ is the distance of the object expressed in geometrical units and $I_{\nu}$ takes the diluted blackbody form \citep[e.g.,][]{1995ApJ...445..780S}.

The redshift factors are calculated as follow:
\begin{equation}
 g=\frac{\nu_{\rm obs}}{\nu_{\rm em}}=\frac{\left.p_{\mu} u^{\mu}_{\rm Observer}\right|_{r=\infty,\theta=\theta_{\rm \rm obs}}}{\left. p_{\mu}u^{\mu}_{\rm Fluid}\right |_{r=r_{\rm em}, \theta=\theta_{\rm em}} }\;,
\end{equation}
where $p_{\mu}$ is the four-momentum of the photon, given by \citet{1968PhRv..174.1559C}
\begin{equation}
p_{\mu}=(-1,\pm \frac{\sqrt R}{\Delta},\pm \Theta^{1/2},\lambda)\;.
\end{equation}
The four-velocity of the fluid is given by \citet{2009ApJ...699..722Y}：
\begin{equation}
u^{\mu}_{\rm Fluid}=(\gamma_{r} \gamma_{\phi} e^{-\nu},\gamma_{r}\beta_{\rm r} e^{-\mu_1},\  0\   ,\gamma_{r}\gamma_{\phi}(\omega e^{-\nu}+\beta_{\phi} e^{-\psi}))\;,
\end{equation}
{
where $e^{ \nu}=(\Sigma \Delta / A)^{1/2}$, $e^{ \psi}=(A \sin^2 \theta/\Sigma)^{1/2}$.}
Thus we have the  expression for redshift factor:
 \begin{equation}
g=\frac{\Sigma^{1/2}\Delta^{1/2}}{A^{1/2}}\frac{1}{\gamma_r \gamma_{\phi}}\frac{1}{1\mp \frac{\beta_r}{\gamma_{\phi}}\frac{R^{1/2}}{A^{1/2}}-\lambda \omega -\lambda \beta_{\phi} \frac{r^2 \Delta^{1/2}}{A}}
 \end{equation}
where ``em'' means the emitting points, and the ``$+$'' sign for the trajectories of the 
photons without $r$-turning point and ``$-$'' sign for
those with $r$-turning point. In Equation(20),
\begin{equation}
R=r^4+(a^2-\lambda^2-Q) r^2+2(Q+(\lambda-a)^2 )r -a^2 Q \;.
\end{equation}
here the definition of $\lambda$ and $Q$ are the same as that of \citet{1973ApJ...183..237C}.

In the ray-tracing calculations, the outer boundary condition of the integration is set at $10^5 r_g$. In the cases when the emitting radius is so large that
the disk solutions do not cover, we use the results from SSD instead, because the differences between the transonic disk model and SSD are indeed insignificant at the large radii.

{
One limitation in our calculation is that the returning irradiation \citep{1995xrbi.nasa...58V,2005ApJS..157..335L} is neglected.
In the inner regions of the accretion disk, due to the strong gravity, the radiation emitted at one part of the accretion disk
may reach other parts, which may result in a modification to the disk structure and the emergent spectra. Certain disk geometry
may enhance such kind of effect \citep{1995xrbi.nasa...58V}. In the case of the thin accretion disk, this effect has moderate
influence (about several percents) on the emergent spectra \citep{2005ApJS..157..335L}. In our case, due to the thick geometry, since the disk is shadowed by itself, the self-irradiation effect may be stronger.
% But we still believe that this is not significant.
As the inclusion of that effect requires significant amount of calculation, for simplicity,
we neglect it in our current work.

Another limitation in our calculation is that the effect of limb-darkening has not been considered. As \citet{2005ApJS..157..335L}
have pointed out, the limb darkening has effect on the emergent spectra, especially when the inclination angle is large,
and the deviation induced by this effect is about $10\%$ at low energy band when the disk is viewed at $85^{\circ}$ \citep{2005ApJS..157..335L}.
In this work, this effect has not been included.
According to the limb-darkening law $I_{\theta} \sim 1/2+3/4 \cos \theta$, our results are still robust at relatively small inclination angles.
}
\begin{figure*}[h!]
\includegraphics[width=0.7 \textwidth]{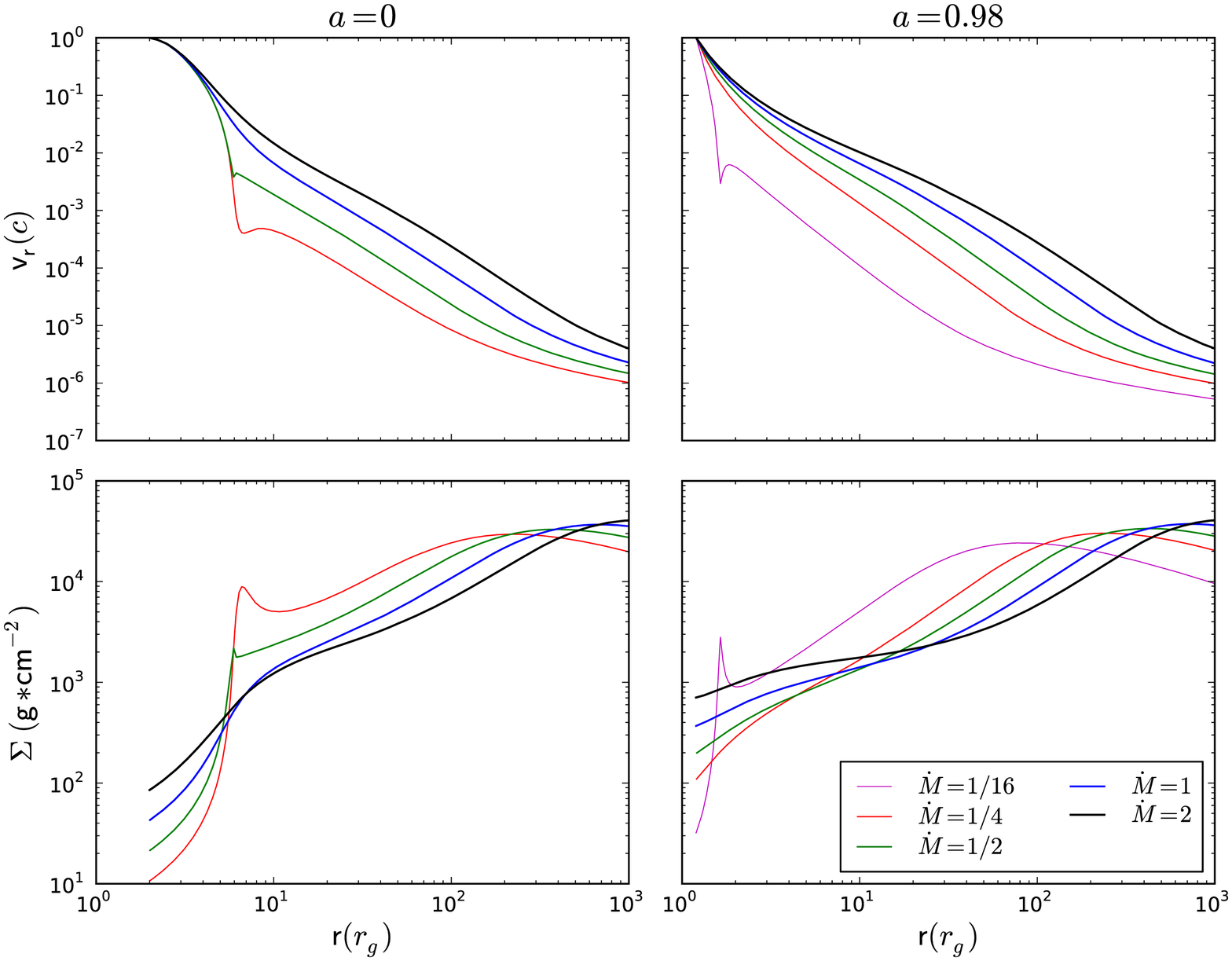}
\caption{\label{vrsigma}Radial velocity and surface density as a function of radii. From thin to thick, the lines stand for the solutions for the
accretion rates $1/16 \dot M_{\rm edd}$, $1/4\dot M_{\rm edd}$, $1/2 \dot M_{\rm edd}$, $1\dot M_{\rm edd}$ and $2 \dot M_{\rm edd}$ respectively. The left panels correspond to Schwarzschild black hole, and the right panels correspond to Kerr black hole with $a=0.98$. The black hole mass is taken to be $10 M_{\odot}$.}
\end{figure*}

\begin{figure*}[h!]
\includegraphics[width=0.7 \textwidth]{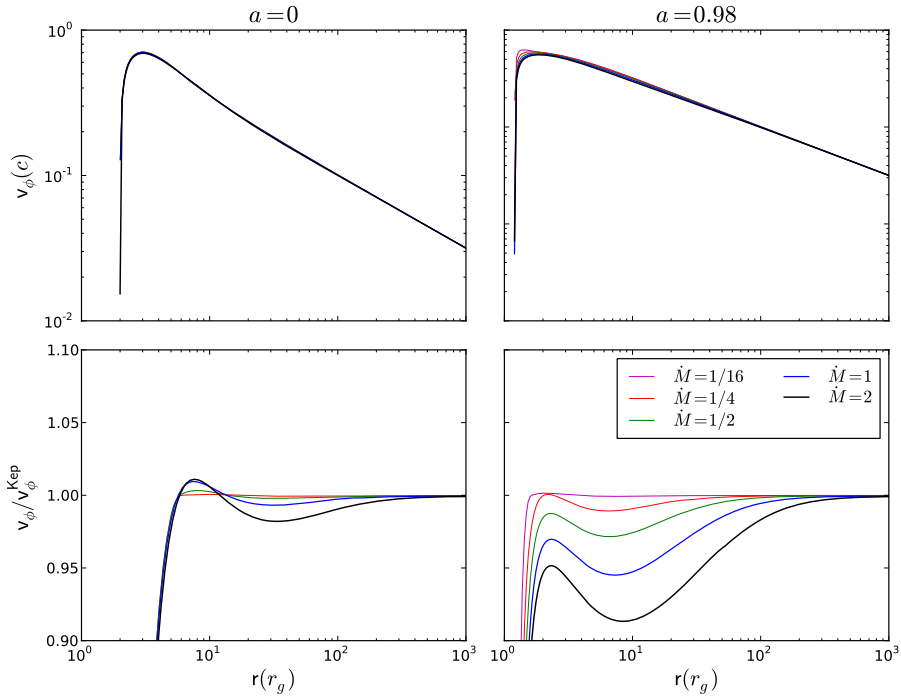}
\caption{\label{vphivphi}Rotation velocity and the ratio of rotation velocity to Keplerian rotation velocity $v_{\phi}/v_{\phi}^{\rm Kep}$ vs. radii. The symbols of the lines are the same as in Figure \ref{vrsigma}. The left panels correspond to Schwarzschild black hole, and the right panels correspond to Kerr black hole with $a=0.98$.}
\end{figure*}

\section{Structure of the Relativistic Disk}

We present our results on the structure of the accretion disk. Throughout this paper, the critical
accretion rate is taken as
\begin{equation}
 \dot M_{\rm edd}=16\frac{L_{\rm edd}}{c^2}\;,
\end{equation}
where the numerical factor comes from the energy generation efficiency of the Schwarzschild black hole. For the sake of comparison,
we adopt the same critical accretion rate for both the Schwarzschild
and Kerr black holes. This is different from the convention used in \citet{2006ApJ...652..518M}. Similar results have already been obtained by \cite{2009ApJS..183..171S}, and one major difference between our work and theirs is that our work is based on a different estimation of the vertical thickness.

In our canonical model, the mass of the black hole is taken to be
$10 M_{\odot}$, while the viscosity coefficient $\alpha$ is set to be 0.1. The dimensionless
accretion rates are taken to be  $1/4$, $1/2$, $1$ and $2$ respectively, ranging from
the sub-Eddington systems to the moderate super-Eddington
ones. A special case for the low accretion rate  $\dot M=1/16$ is also discussed when  $a=0.98$.

\subsection{Disk Dynamics}
\label{secst}

\begin{figure*}[h!]
\includegraphics[width=0.7 \textwidth]{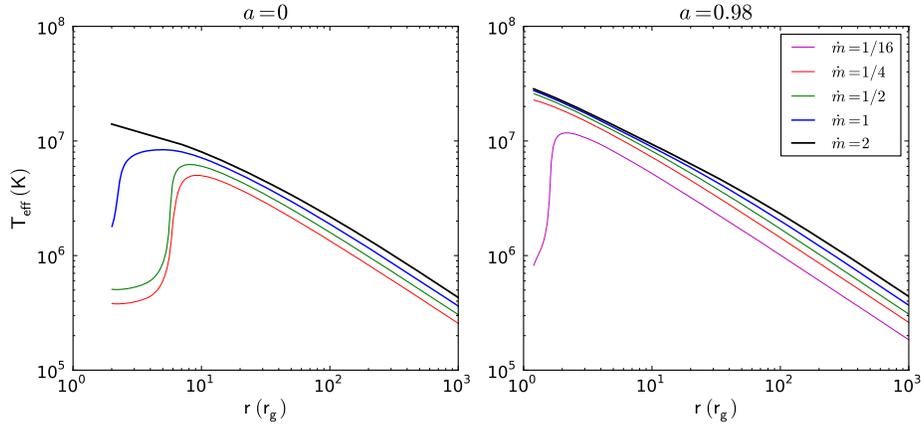}
\caption{\label{Teff}Effective temperature $T_{\rm eff}$ vs. radii $r$. The symbol of the lines are 
the same as in Figure \ref{vrsigma}. The left panels correspond to the Schwarzschild black hole, and the right panels correspond to the Kerr black hole with $a=0.98$.}
\end{figure*}

Figures \ref{vrsigma}-\ref{Teff} show the dynamical properties of our accretion disks.
The upper panels in Figure \ref{vrsigma} show the radial velocity as a function of radii, with the different accretion rates.
Because we use fully relativistic equations, near to the event horizon of the black hole, as expected, the radial velocity is very close to the speed of light.
At the small radii, for the same accretion rate, the higher the spin, the smaller the radial velocity.
At the large radii,
the radial velocity of the accretion flow is relatively small, and it is insensitive to the black hole spin.
This is because that at the large radii, the disk structure resembles that of SSD.
With the increase of the accretion rate, the radial velocity becomes larger.

The lower panels of Figure \ref{vrsigma} show the radial distribution of the surface density
under the different accretion rates. Generally speaking, at the small radii, the surface
density is small.
These results can be simply understood
in terms of the non-relativistic form of the matter conservation:
\[
\dot M =2 \pi r \Sigma v_{\rm r} \;,
\]
where $\Sigma$ is the surface density of the accretion disks, and $v_{\rm r}$
is the radial velocity (the relativistic one is similar but with an additional corrections
related to the radial Lorentz factor $\gamma_{\rm r}$ and the
black hole spin $a$). Since the accretion
disks around Schwarzschild black holes tend to have higher radial
velocity, their surface density at given radius is smaller. Another
noticeable feature is that the density hop occurs when $\dot M=0.25 \dot
M_{\rm edd}$, $a=0$, which is accompanied by the velocity decrease and
re-increase in the upper panels. This peculiar behavior is not
shown in \citet{2003MNRAS.338.1013S} since they did not
explore the disk solution at such a low accretion rate, but can be
found in the previous papers solving the non-relativistic slim disk
structure \citep{2004ApJ...614..101C} and a recent relativistic one
\citep[][with explanation]{2009ApJS..183..171S}.

Upper panels of Figure \ref{vphivphi} show the rotation velocity of the accretion flow.
The rotation velocity is insensitive to the accretion rate,
 especially at the outer region. To show how the fluid motion deviates from the Keplerian motion,
the ratio of the rotational velocity to the rotational velocity of the Keplerian motion $v_{\phi}/v_{\phi}^{\rm Kep}$
are shown in the lower panels of Figure \ref{vphivphi}. When the accretion rate is low,
 the rotation velocity ratio is very close to unity, which  means the assumptions made in SSD are accurate.
 When the accretion rate is relatively high, the rotation of the fluid becomes sub-Keplerian. In some regions, the rotation is super-Keplerian.
Although there exists some differences, the deviation is no more than $10\%$ at several gravitational radii,
which means the assumption of the Keplerian motion is still valid at the most of the regions of the accretion disks \citep[see e.g.][]{2000ApJS..130..463I}.

Figuere \ref{Teff} shows the effective temperature $T_{\rm eff}$ as a function of radii. Here, $T_{\rm eff}$ is
defined as $T_{\rm eff}=( F_{\rm rad}^-/\sigma
)^{1/4}$, where $F_{\rm rad}^-$ is the radiation flux per unit
area and $\sigma$ is the Stefan--Boltzmann constant. Similar to the above figures, the differences between the
Schwarzschild and the Kerr black holes are all very small at large radii.
With the increase of the accretion rate, the temperature increases.
When the accretion rate is high enough, the temperature is still high even inside the ISCO, 
which suggests that, contrary to the assumptions made in the SSD model, the matter inside the ISCO is not cold.

\subsection{Disk Thickness}
\begin{figure*}[h!]
\includegraphics[width= 0.7 \textwidth]{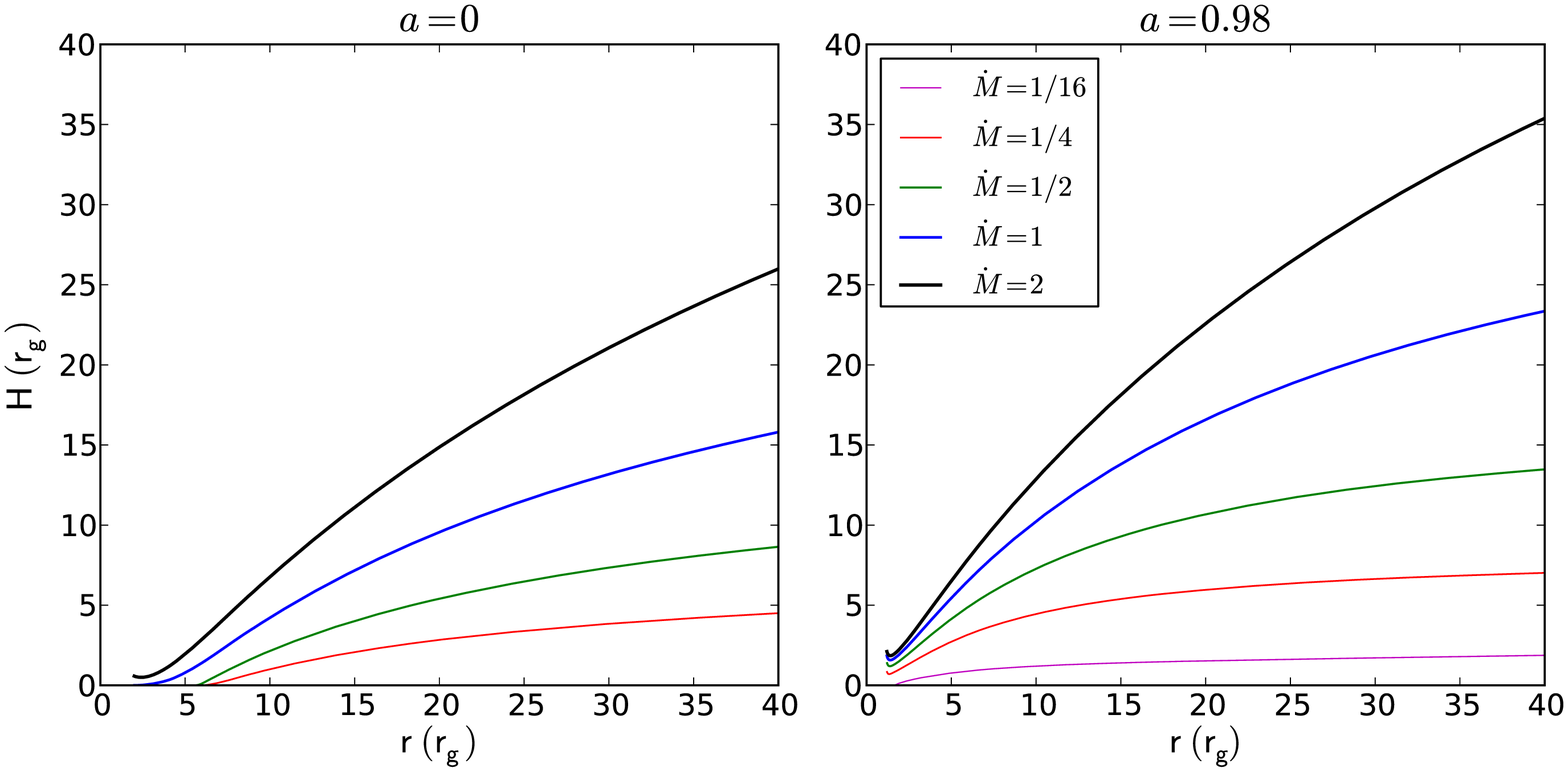}
\caption{\label{H}Vertical thickness as function of radius. The symbol of the lines are the same as in Figure \ref{vrsigma}. The left panels correspond to the
Schwarzschild black hole, and the right panels correspond to the Kerr black hole with $a=0.98$.}
\label{fig:line}
\end{figure*}
Figure \ref{H} shows the thickness of the disk vs. the radii.
In order to give a instinctive picture of the
 disk thickness, we use linear coordinate, and show only the results in the inner part of the accretion disk. This thickness
 profile is thus correspond to the surface of the accretion disk. The imprints of ISCO appear in the disk thickness profile.
  This feature is particularly prominent at low accretion rate, where the disk surface agrees with
   the simulation carried out by \citet{2002ApJ...566..164H} and \citet{2008ApJ...675.1048R}.
% , and seems to follow the equi-potential surface as suggested by Paczynsky  \citep{2009ASPC..403...29A}.
When the accretion rate is high enough, the imprints of ISCO disappear, and the disk shows a continuous profile down to the event horizon of the black hole.

However, in some cases, the estimation of thickness may be incorrect. Since the estimation of the thickness is based on the vertical $hydrostatic$
   equilibrium, it depends on the assumption:
\begin{equation}\label{eq25}
a_{\rm z}^{\rm Fluid}<g_{\rm z}\;,
\end{equation}
which means the acceleration in the vertical direction could never exceed the acceleration provided by the vertical gravitational force (or that the disk has time to relax to the hydrostatic equilibrium).
We checked our solutions using this formula, and found that in some cases such as $\dot M=0.25$, $a=0$ Equation (\ref{eq25}) is not satisfied.
Thus the disk thickness may be under-estimated at the regions near ISCO, in some cases.
% \subsection{Vertically Integrated Stress}
%
% Fig . \ref{stress} shows the integrated stress as a function of the radii $r$. Similar to previous discussions, the imprints of
% ISCO also appear. When the  accretion rate increases, the imprints of ISCO tend to disappear.
%
% \begin{figure*}[h!]
% \includegraphics[width=14cm]{stress.eps}
% \caption{\label{stress}Vertical integrated stress as function of $r$. The symbol of the lines are the same as in Figure \ref{H}.}
% \label{fig:line}
% \end{figure*}
%\subsection{Energy Conservation and Dissipation Inside ISCO}
\subsection{Energy Dissipation Inside ISCO}

\label{secend}

% \begin{figure*}[h!]
% \includegraphics[width=14cm]{Qvis.eps}
% \caption{\label{vis} Viscous heating rate as a function of $r$. The symbol of the lines are the same as in Figure \ref{H}.
% }
% \label{fig:line}
% \end{figure*}

\begin{figure*}[h!]
\includegraphics[width=0.7 \textwidth]{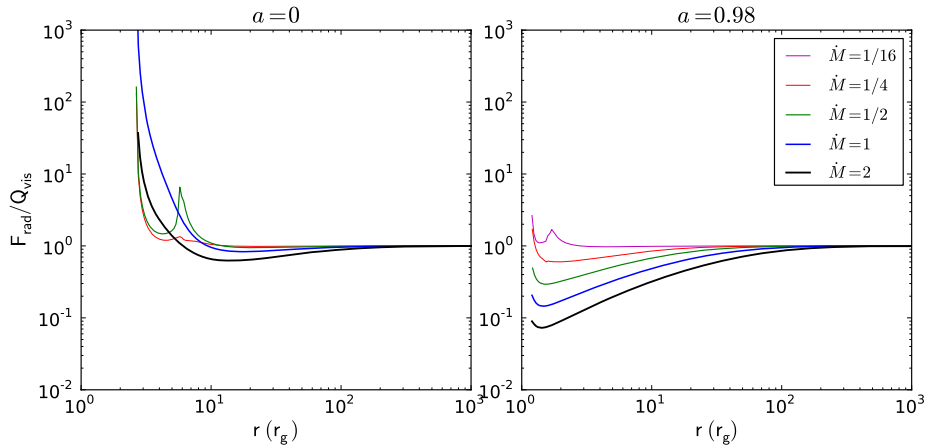}
\caption{\label{frac}$\frac{F_{rad}^-}{Q_{vis}^+}$ as function of $r$. The symbol of the lines are the same as in Figure \ref{H}.
}
\label{fig:line}
\end{figure*}

% In Figure \ref{vis}-
In Figure \ref{frac}, we present our results on the global energy balance of the accretion disks.
The energy conservation in the accretion disks takes the simple form as follows:
\begin{equation}
 Q_{\rm{vis}}^+=F_{\rm{rad}}^-+Q_{\rm{adv}}^-\;,
\end{equation}
where $Q_{\rm vis}^+$ is the viscous energy generation rate, $F_{\rm rad}^-$ is the cooling term due to radiation from the accretion disk,
and $Q_{\rm adv}^-$ is the effective cooling induced by the heat advection. If $Q_{\rm vis}^+\sim F_{\rm rad}^-$, the disk is very similar to
the standard accretion disk since the energy generated by the viscous processes is radiated locally. If $Q_{\rm vis}^+ \sim Q_{\rm adv}^-$, most of the heat generated is trapped inside the accretion disk.

% Figure \ref{vis} shows the viscous heating rate as function of radii. When the accretion rate is low enough, the dissipation drops significantly at ISCO. When the accretion rate is high enough, substantial amount of dissipation happens inside the ISCO.

Figure \ref{frac} shows the ratio of the emitted radiation from the disk surface to the total heat generated by the viscous processes.
At about 20 $r_{\rm g}$,
  the heat advection works, as a result, the ratio is less than $1$. However, in the inner region, in many cases,
   $F_{\rm rad}^-$ dominates over $Q_{\rm vis}^+$, which means that substantial amount of energy is still being radiated away. The source of the energy mainly
    comes from the energy trapped in the region outside the ISCO.
Thus our results suggest that the radiation trapped outside the ISCO can be allowed to escape at the regions close to the black hole.
This kind of mechanism is new, since different from the advective cooling, in some cases, the effect of advection will heat the disk up,
resulting in ``advective heating''.

Besides from getting self-consistent disk solution, we find that when the accretion rate is low
enough such as $\dot M \sim 1/4$ for $a=0$ and $\dot M \sim 1/16$ for $a=0.98$, the SSD model is valid.

\begin{figure*}[here!]
\centerline{\rotatebox{0}{\includegraphics[width=0.85\textwidth]{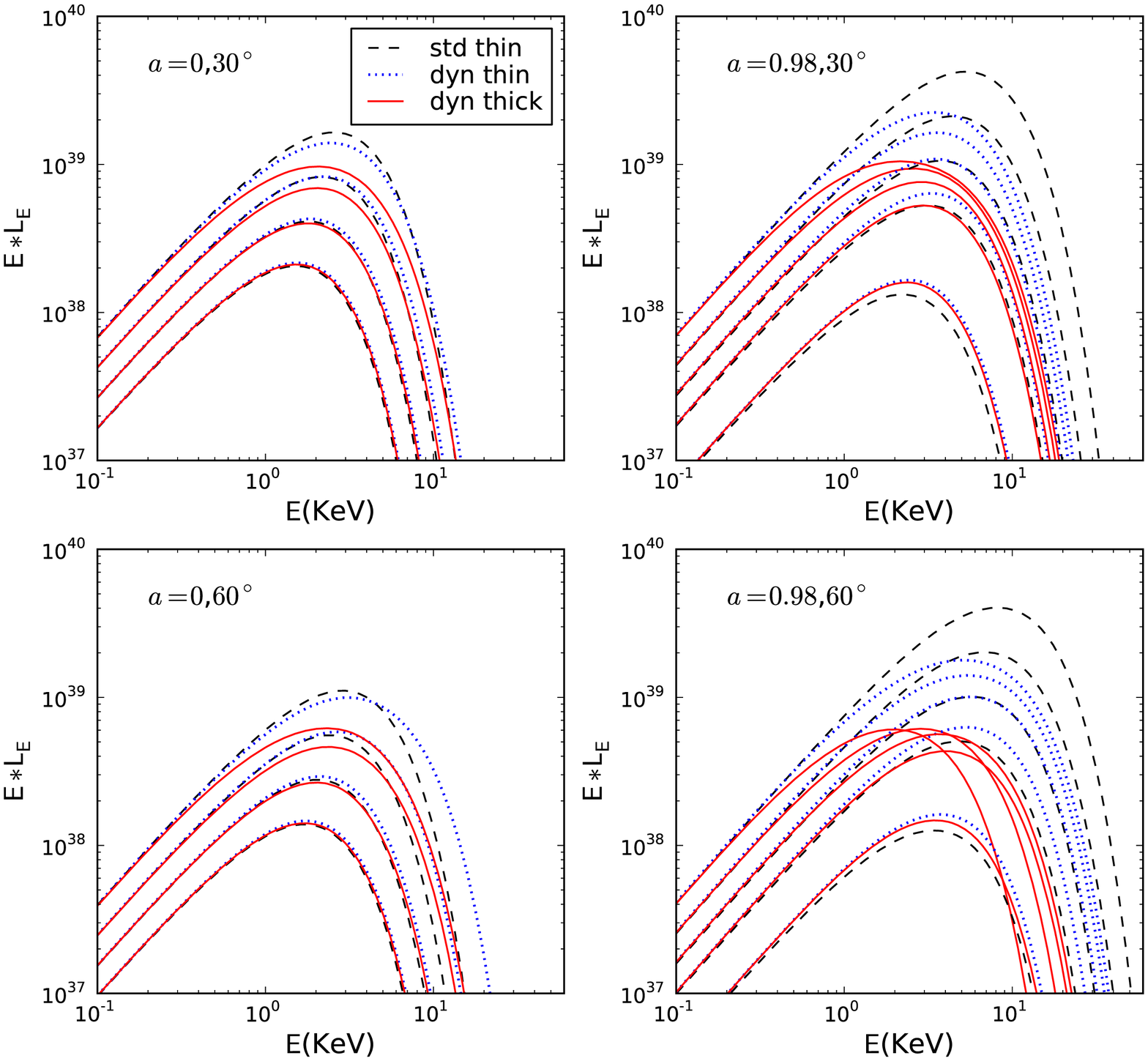}}}
 \caption{\label{ag}\small\label{spectral}
The emergent spectra.
The upper panels show the spectra when
the disk is viewed at $30^{\circ}$, while the lower panels show the
spectra at $60^{\circ}$. The left panels show the spectra from the
non-spinning black holes ($a=0$), the right panels show the
spectra from the spinning black holes ($a=0.98$). The black dashed lines
correspond to the spectra from the relativistic standard disks (case A),
the blue dotted lines correspond to the spectra form the relativistic
slim accretion disks with no thickness (case B), and the red solid
lines correspond to the spectra from the relativistic slim accretion
disks with the inclusion of thickness effect (case C). The accretion
rate takes the values: $1/4 M_{\rm edd}$, $1/2 M_{\rm edd}$, $1 M_{\rm edd}$, $2
M_{\rm edd}$(left panels, $a=0$) and $1/16 M_{\rm edd}$, $1/4 M_{\rm edd}$, $1/2 M_{\rm edd}$, $1 M_{\rm edd}$, $2
M_{\rm edd}$(right panels, $a=0.98$) respectively. The black hole mass is taken to be $10 M_{\odot}$.
%Both the disk dynamics and disk thickness have significant effect on the emergent spectra.
 }
\end{figure*}
\section{Emergent Spectra}
\label{respec}

We present the resulting emergent spectra based on the ray-tracing method.
We assume that the local emission of the disk takes the diluted blackbody form with the spectra hardening
factor equals 1.7 \citep{1992MNRAS.258..189R,1993ApJ...419...78S,1995ApJ...445..780S}, which is actually
dependent on disk luminosity as shown in \citet{2005ApJ...621..372D} and \citet{2008ApJ...683..389D}.

To show the effect of the photon trapping and the disk thickness, we consider three cases. In the first case (case A),
we calculate the spectra from the relativistic standard accretion disks. In this case, our spectra are completely
identical to the spectra generated by KERRBB with exactly the same parameters. In the second case (case B),
we calculate the spectra from the relativistic slim accretion disks, assuming that the disk is still geometrically thin.
 This is for the purpose of showing the effect of the heat advection, the sub-Keplerian motion, the radial plunging and the radiation from the plunging region.
 In the third case (case C), based on case B, we consider the effect of disk thickness to
 show the disk self-shadowing.
% \subsection{Numerical Results}

The emergent spectra are shown in Figure \ref{spectral}. The differences between case A (black
solid line) and case B (blue dotted line) are due to the combination
of the effects of the photon trapping, the radiation from the plunging region and
the motion of the fluid. The effect of the motion of the fluid on the emergent
spectra is determined mainly by the Lorentz factors, $\gamma_{r}$,
$\gamma_{\phi}$. At the $\phi$ direction, the velocity differences
between the slim disk model and the thin, Keplerian SSD model is relatively
small, thus the effect is not significant. At the $r$ direction, the
radial velocity works. Since matter plunges into the
black hole, in most cases, this will introduce a net redshift. At
some rare cases where the photon trajectories have $r$ turning points
\citep*[e.g][]{1998NewA....3..647C}, the radial velocity field will
introduce net blue--shift. But this blue shift is usually
unimportant since this kind of trajectory is rare. The radial velocity effect is significant only near or inside the ISCO.

The other important effect is the photon trapping. Since
the previous works focus on the application of slim disk model in the highly
accreting systems, the effects of the photon trapping in their cases are thus to bring down
the luminosity, and highly attenuate the flux at the high energy band.
However, for the relatively low accretion rate, the photon
trapping behaves quite differently. In fact, contrary to the
previous view that photons are trapped and not allowed to radiate
away, substantial amount of trapped photons is allowed to re-radiate in
the plunging region where the energy dissipation is relatively insignificant. This
effect will in fact distort the spectrum and hence affect the spin
measurement. This phenomenon is common for both the non-rotating and rotating black hole
systems when the accretion rate is moderate.

The differences between case B and case C are due to the inclusion of the thickness of the disk in our calculations.
It is clearly seen that the inclusion of the effect of the disk thickness will bring down the flux, especially at high energy
band, and is more prominent at the higher accretion rate, higher
inclination angle and larger spin. When the accretion rate increases,
the disk thickness increases, and the self-shadowing happens at a relatively small inclination angle. For
the Kerr black hole, since the energy generation rate is higher, the
disk self-shadowing effect is more prominent.

Since the differences between case A and case B are what referred to as the photon trapping effect,
the differences between case B and case C are due to the effect of the disk self-shadowing, which has never been
worked out before. In fact, when the accretion rate is high, the
effect of the disk self-shadowing is more significant than that of the photon-trapping effect alone.
Thus, in order to study the spectra from the super-critical accretion disks,
we have to consider those two effects simultaneously.

{
According to our results, for the disk around a Kerr black hole with $a=0.98$, if the disk inclination is
$60^\circ$, at about $0.23$ of Eddington luminosity, the effects of the
heat advection and disk self-shadowing begin to be important. Although in our case (the bottommost set of curves in Figure \ref{spectral}),
the effects of the heat advection and disk self-shadowing tend to cancel each other, our results set a critical luminosity beyond which
the relativistic standard accretion disk model is no longer applicable for the spectra fitting.
}

\section{Angular Dependence of the Luminosity}
\label{angulardep}

\begin{figure*}[here!]
\centerline{\rotatebox{0}{\includegraphics[width=0.85\textwidth]{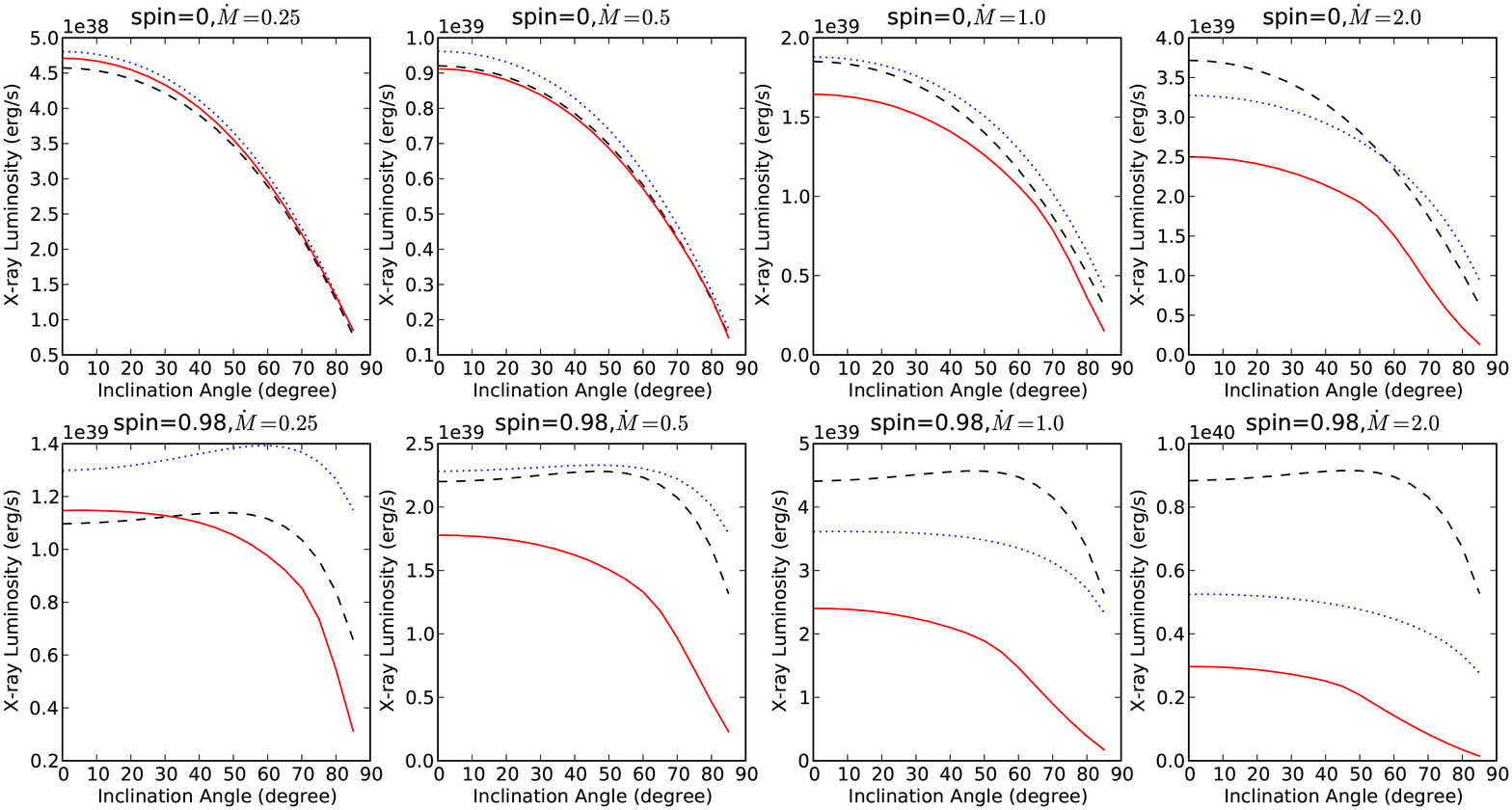}}}
 \caption{\label{ag}\small\label{dep1}
The angular dependence of the observed X-ray luminosity. As in Figure \ref{spectral}, the (black) dashed lines stand for
the relativistic standard accretion disk, the (blue) dotted line stands for the dynamical thin accretion disk,
and the (red) solid lines stand for the dynamical thick accretion disk.
Different panels correspond to different black hole spin and accretion rate. The black hole mass is taken to be $10 M_{\odot}$.
 }
\end{figure*}

\begin{figure*}[here!]
\centerline{\rotatebox{0}{\includegraphics[width=0.85\textwidth]{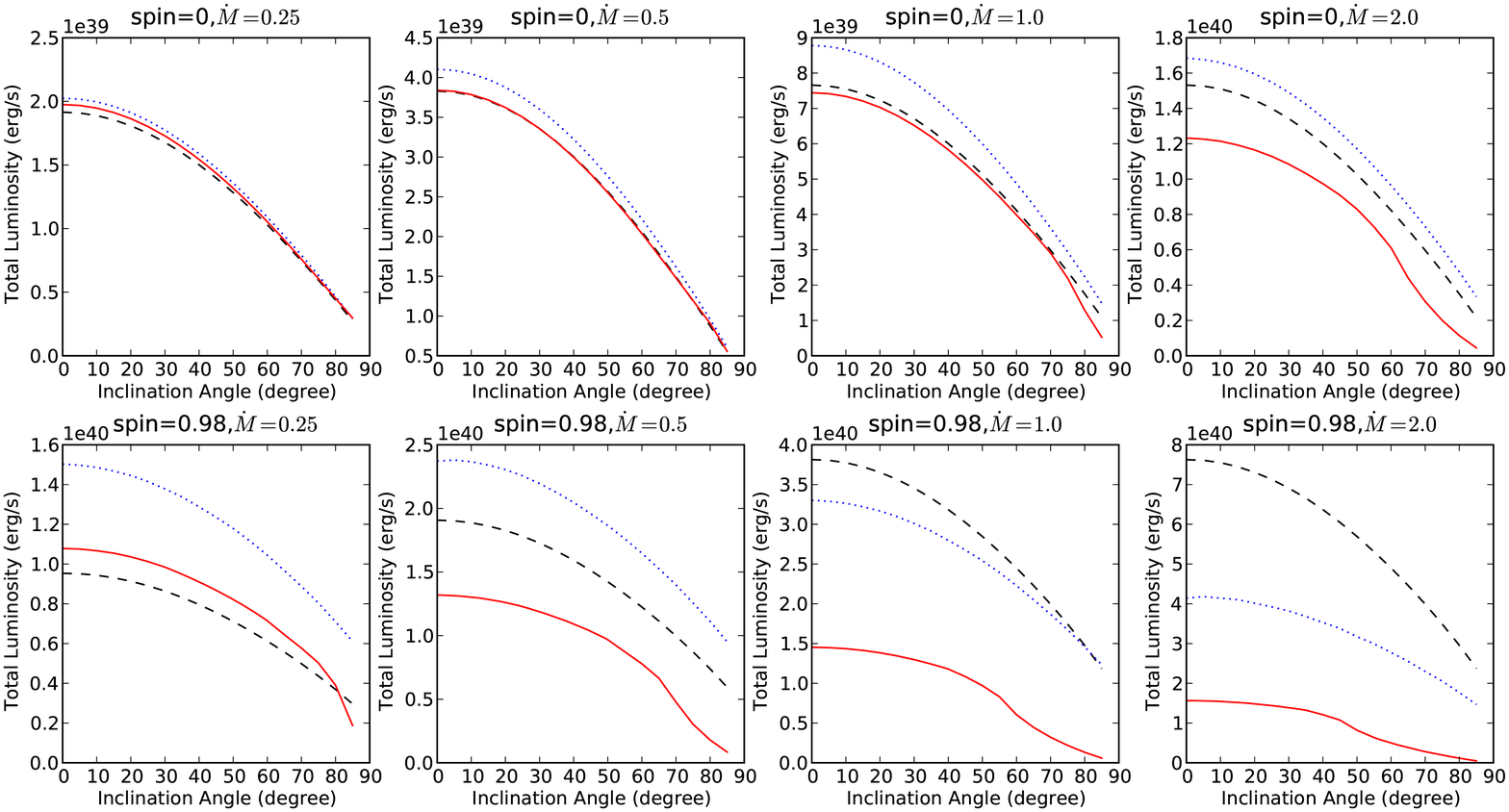}}}
 \caption{\label{ag}\small\label{dep2}
The angular dependence of the observed total luminosity. The symbol of the lines are the same as in Figure \ref{dep1}.
 }
\end{figure*}

We explore the angular dependence of the disk luminosity. In the case of a thin slab of the stationary gas,
the angular dependence of emergent luminosity  should be $L \sim \cos \theta $. In the case of an accretion disk,
the angular dependence of the disk luminosity $L(\theta)$ is also influenced by several other effects, including the disk geometry and the disk rotation.
In Figures \ref{dep1}-\ref{dep2}, we show both the luminosity in the X-ray band ($0.1\sim100 \rm KeV$), which is of observational interest, and the total luminosity.

Figure \ref{dep1} shows the luminosity in the X-ray band as a function of the inclination. The upper panels show the results for a non-spinning black hole,
and the lower panels show the spectra for a spinning black hole with  $a=0.98$. For the standard accretion disk model, the angular dependence of the luminosity
deviates from $L\sim \cos \theta$ due to the Doppler shift induced by motion of the fluid. For the disk around a Schwarzschild black hole, the effect of the
Doppler shift tends to flatten the profile, which means that the radiation is more weakly beamed than that of a single slab, for the disk
around a Kerr black hole, this effect is so strong that the angle at which luminosity peaks is not $0^{\circ}$.

The dotted lines in the subpanels of Figure \ref{dep1} show the angular dependence of the X-ray luminosity for a dynamical thin accretion disk
which is different from the standard accretion disk since its velocity field is different and the photon trapping effect is taken into account.
For the low accretion rate, as we have discussed, the effect of photon trapping is complicated. For the high-enough accretion rate,
the effect of heat advection will highly attenuate the total luminosity.

The solid lines in the subpanels of Figure \ref{dep1} show the angular dependence of X-ray luminosity for a dynamical thick accretion disk.
Compared to the blue dotted lines, the effect of disk thickness is taken into account. In all cases, the inclusion of disk thickness tends to
attenuate the radiation, even in the face-on case, since light is bend by the strong gravity. Another difference caused by the disk thickness
is a mild beaming of radiation \citep[e.g.][]{2006MNRAS.372.1208H}, which means that the disk is more luminous when viewed face-on than viewed edge-on.

The angular dependence of the total luminosity is shown in Figure \ref{dep2}. The trend is similar to the previous figure.
However, the effect of the motion of the fluid is relatively weak. The reason is that in calculating the X-ray luminosity,
the low-energy photons can be Doppler boosted into the X-ray band and contributed to the X-ray luminosity.
However, that effect will not significantly influence the total luminosity.

The mass of the black hole in our model is taken to be $10 M_{\odot}$. Since the disk thickness in the inner
regions of the accretion disks is dependent on the accretion rate but independent of black hole mass \citep[][]{1998bhad.conf.....K,2006ApJ...648..523W},
for a given $\theta_{\rm crit}$, the effect of self-shadowing is only a function of the black hole accretion rate.
Given the same Eddington-scaled luminosity, since the other physical parameters such as $T$ (the temperature or equivalently the characteristic frequency) and
$L_{\rm tot}$ (the normalization of the flux) scale as
\begin{equation}
 T \sim M^{-\frac{1}{4}}
\end{equation}
\begin{equation}
 L_{\rm tot} \sim M \;,
\end{equation}
our results can be scaled to a considerable range of parameters,
and then can be used to fit the observational results of the supermassive black holes.

% One place to test our results is the X-ray binary. In some cases, we can measure the inclination of the system \citep[e.g.][]{1999MNRAS.304..865F}(which is believed to be the inclination of the accretion disk). Thus by tracing the evolution of such systems in which accretion rate changes, we can track the evolution of the accretion disk \citep[e.g.][]{2004ApJ...601..428K,2004MNRAS.353..980K,2007A&ARv..15....1D}, and test if the accretion disk is shadowed by itself.

\section{Implications for Spin Measurement} \label{reimp}
\begin{table}[h!]
\caption{Parameters of KERRBB} \label{tab_params}
\begin{tabular}{llll}
\tableline\tableline
Parameters Name & Description & Value & Freeze\\
\tableline
$\eta$ & Inner torque & 0 & Yes\\
$M_{\rm bh}$ & Mass of black hole & 10 & Yes \\
$\theta_{\rm obs}$ (degree) & Disk inclination angle & $30$, $60$\footnotemark& Yes \\
$D$ (kpc) & Distance & $10$ & Yes \\
rflag & Self--radiation & $0$ & ... \\
lflag & Limb--darkening & $0$ & ... \\
$f_{\rm col}$ & Spectral hardening factor & $1.7$ & Yes\\
$K$ & Normalization & $1$ & Yes \\

$\dot M$ & Accretion rate & ... & No \\
$a$     & Spin  & ... &   No \\
\tableline
\end{tabular}
\footnotetext{The values used in the fittings are the same as the value used in producing the spectra}
%\footnotetext{Fitting parameter}
\end{table}

We show how the different SEDs in different models affect the measurement of the black hole spin. To achieve this, we create ``fake''
 data sets from our spectra by convolving with a response matrix of XMM-NEWTON using XSPEC \citep{1996ASPC..101...17A} version 12.5. The ``faked'' data sets are then fitted
 with the KERRBB model in XSPEC.

The KERRBB model is a publicly available XSPEC model that can be used to calculate the multicolor disk (MCD) spectra of a relativistic, thin accretion disk
around a Kerr black hole. The model settings in our fittings are
summarized in Table \ref{tab_params}. Our fitting strategy is similar to that of \citet{2006ApJ...636L.113S}.
The differences are that we turn off the limb-darkening and self-radiation, and we always set the spectral hardening factor $f_{\rm col}$ to be $1.7$.
This is because we require the fittings consistent with the assumptions in our model.

The fittings are carried out by making our calculated spectra into a table model,
 then the faked data can be easily generated using the $fakeit$ command. The faked
 spectra are then fitted to model KERRBB with the procedure described above.
 The fittings are robust in the sense that the different initial values of black hole spin $a$ and the accretion rate $\dot M$
 lead to almost the same results.

\begin{figure}[h!]
%\hfill
\includegraphics[angle=0,width=0.35\textwidth]{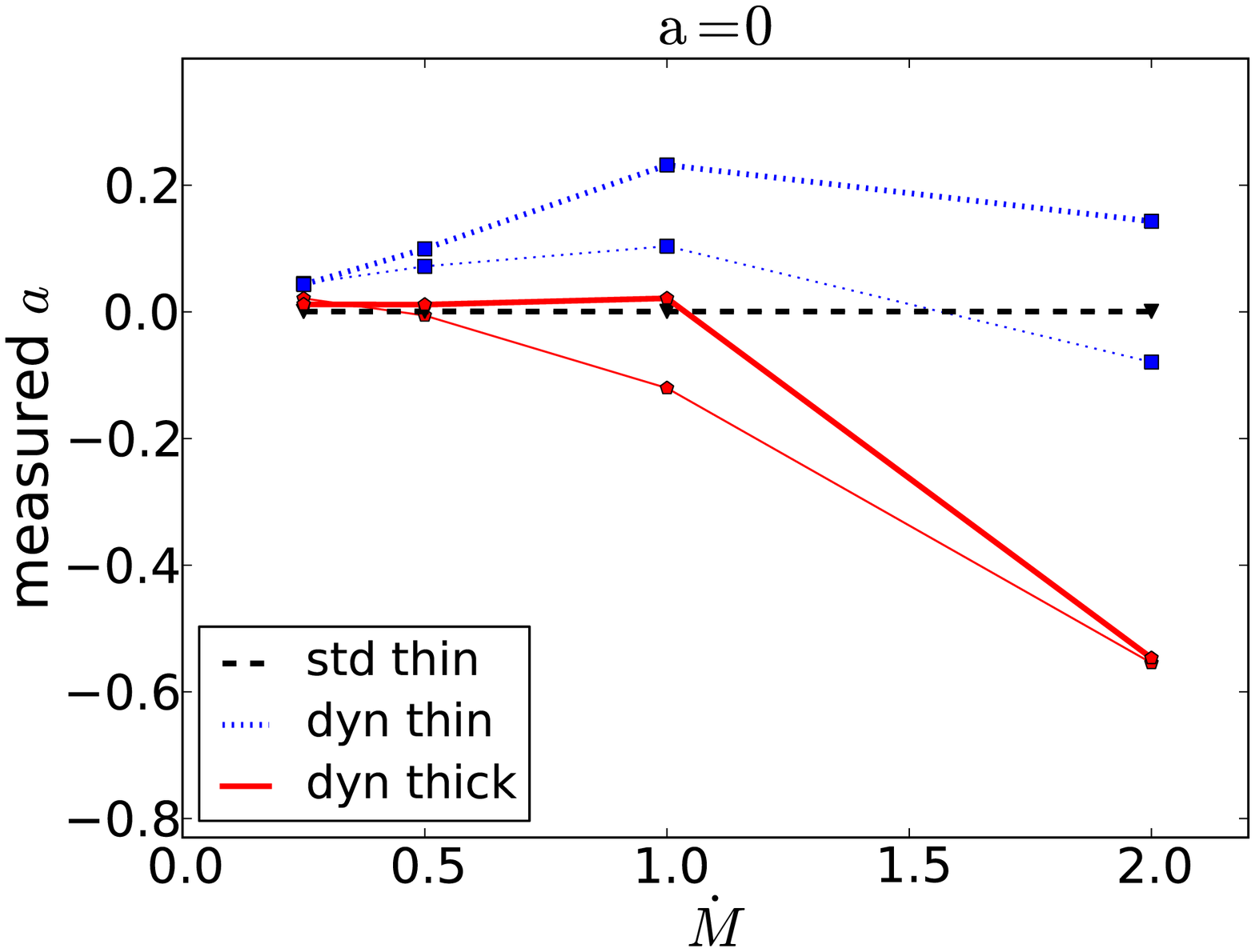}
\hfill
\includegraphics[angle=0,width=0.35\textwidth]{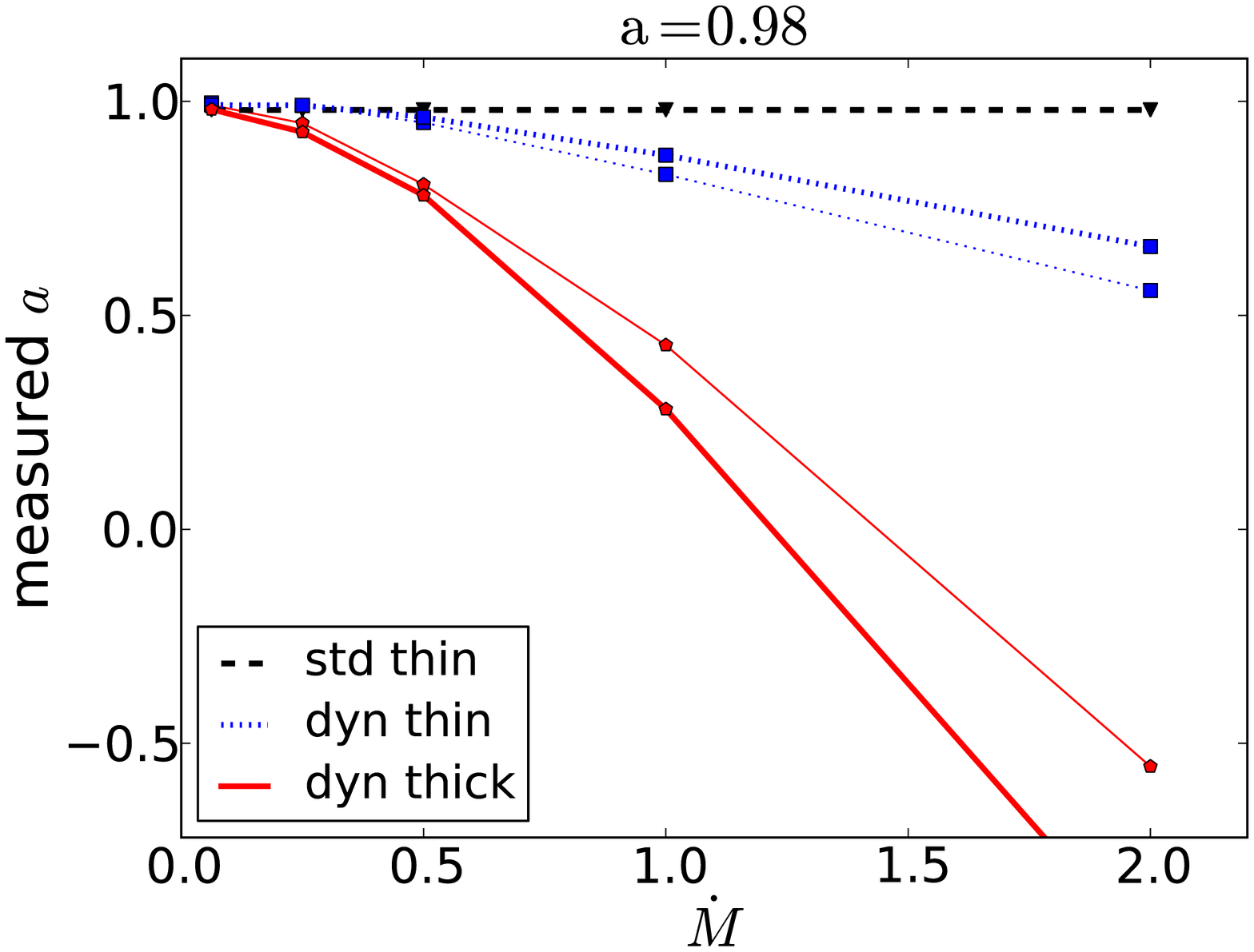}
\hfill

\caption{\label{fitting}Measured spin for different models. The upper
panel shows the result for a non-spinning black hole with $a=0$, and the
lower panel shows the result for a spinning black hole with $a=0.98$. The
dashed lines correspond to the thin Keplerian case (case A), the dotted
lines correspond to the slim accretion disks with no thickness (case B), and
the solid lines correspond to slim accretion disks with thickness
effect (case C). Different line thickness represent different viewing
angles: Thin lines for $\theta=30^{\circ}$, and thick lines for
$\theta=60^{\circ}$.} \label{fig:line}
\end{figure}
% \begin{figure}[here!]
% %\hfill
% \includegraphics[angle=0,width=0.35\textwidth]{mashow.eps}
% \hfill
% \includegraphics[angle=0,width=0.35\textwidth]{mbshow.eps}
% \hfill
%
% \caption{\label{mfitting}Measured accretion rate for different
% models. The symbol of lines are the same as in Figure \ref{fitting}}
% \end{figure}
Figure \ref{fitting} shows the measured spin. For the
moderate accretion rate, the effect of the heat advection alone leads to
the overestimation of black hole spin, while for the high-enough accretion
rate, this usually leads to the underestimation of the black hole
spin. This is true for both the Schwarzschild and Kerr cases, and for the
Kerr case these phenomena occur at a relatively lower accretion
rate. The effect of the disk thickness will always lead to the under-estimation
of black hole spin and is very significant when the accretion rate,
the inclination or the spin, is large. Combining these two effects, in
most cases, we generally underestimate the black hole spin. Only when the
physical parameters are appropriate, we can obtain a overestimation of the black hole spin.

% Figure \ref{mfitting} shows the measured accretion rate. {\bf Although in Figure \ref{spectral} the predicted luminosity for our model and SSD differs at about $50\%$, since the SED is plotted in $\nu L_{\nu}$ plane while the fittings are carried out by considering the photon number only, a significant deviation in $\nu L_{\nu}$ plane will only introduce a small deviation in the photon counting. Thus the measured accretion rate is not significantly
% influenced by the heat advection and the disk self-shadowing effects. Thus our results suggest
% that the measurement of accretion rate through spectral fitting is relatively reliable. }

\section{Discussions}
{
\label{disc}
\subsection{Shadowing Effect Enhanced by the Strong Gravity}
\begin{figure}[here!]
 \includegraphics[width=0.4 \textwidth]{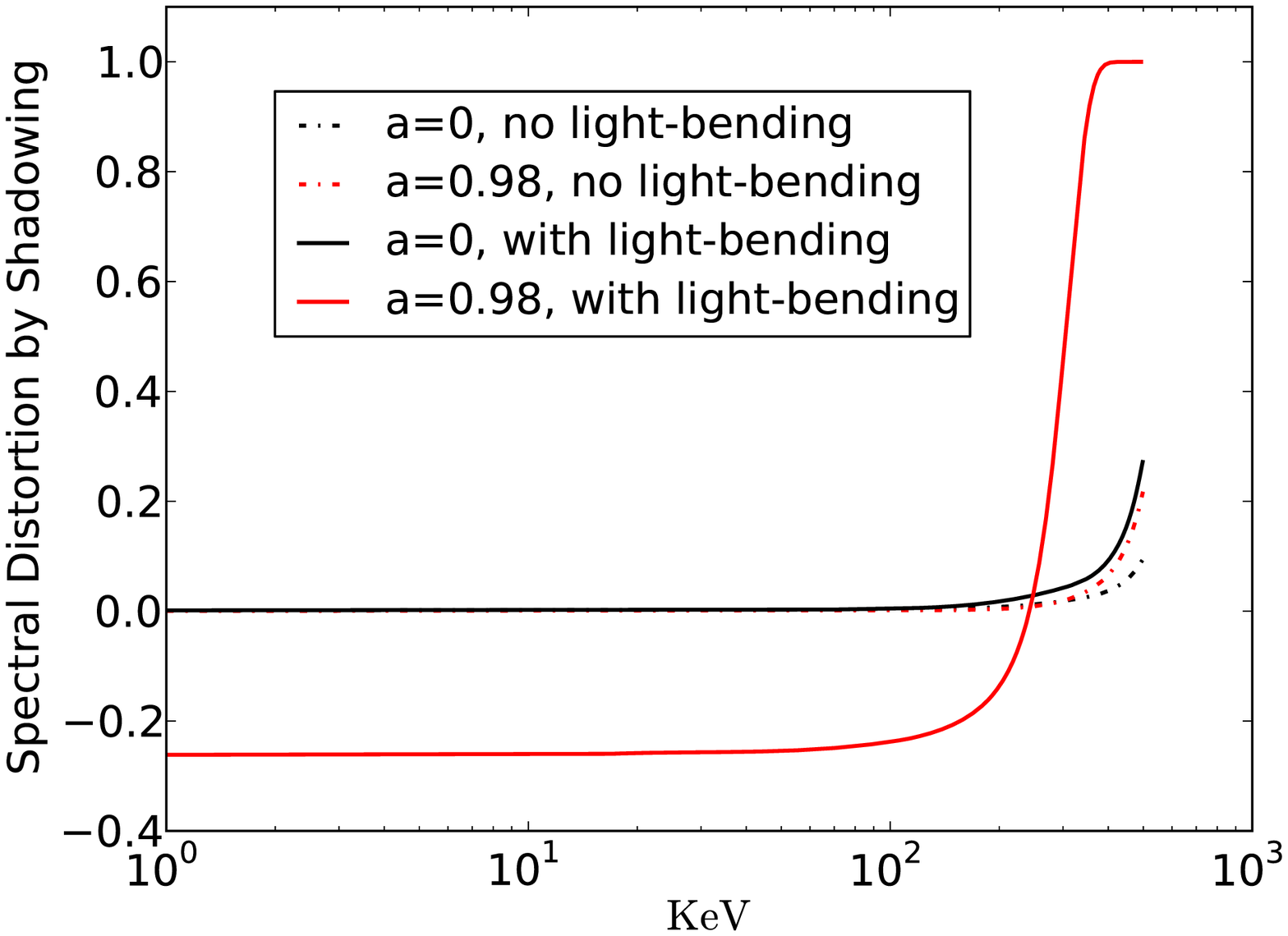}
 \caption{\label{distortion}
Spectral distortion (defined by Equation \ref{eqdist}) as a function of the photon energy. The solid lines standard for the results
with the effect of the light bending, while the dotted lines for the results without the light-bending effect.
The red lines stand for the Kerr black hole with $a=0.98$, and the black lines for the black hole with $a=0$.
In all the cases, the disk luminosity is the same.  }
\end{figure}
It is of theoretical interest to investigate what influences the effect of the self-shadowing. 
Without the light-bending effect, the self-shadowing effect is dependent on the maximum of $H/r$, 
which is mainly dependent on the black hole accretion rate \citep{2006ApJ...652..518M}. 
However, it remains unclear how the shadowing effect influenced by the strong gravity.

To gain physical insight into this question, we choose two systems with exactly the same luminosity 
but the different spin (one with $\dot M=1$ and $a=0.974$, another with $\dot M$=4 and $a=0$), and 
calculated the spectra. For both systems, we calculate the spectra 
both with and without the effect of light-bending. In the latter case, for self-consistency, 
the redshift is evaluated as
\begin{equation}
g=\frac{\Delta^{1/2} r}{\gamma_{r}\gamma_{\phi}A^{1/2}}\frac{1}{1\pm \frac{\beta_{\rm r} \sqrt{R}}{\gamma_{\phi}A^{1/2}}-\Omega \lambda}\;.
\end{equation}
After that, to quantify the effect of disk self-shadowing, we compare the SED calculated with disk thickness and the SED calculated without disk thickness by defining the spectral distortion as function of frequency as

\begin{equation}\label{eqdist}
{\rm Distortion(\nu)}=\frac{\rm SED_{\rm not\ shadowed}-\rm SED_{\rm shadowed}}{\rm SED_{\rm not\ shadowed}+SED_{\rm shadowed}}\;,
\end{equation}
where the $\rm SED_{\rm not\ shadowed}$ denotes the SED calculated by assuming slim disk with no thickness, and $\rm SED_{\rm shadowed}$ denotes the SED calculated including the disk thickness.

The spectral distortion for different spin and different assumptions about light trajectories is plotted in Figure \ref{distortion}. For both the Schwarzschild and the Kerr black hole, including the light bending effect brings about significant distortion to the emergent spectra, and this distortion generally appears as attenuation of the flux at high energy part. This kind of distortion is more significant for Kerr black hole, because the light trajectories are more severely influenced by the strong gravity.

%
% To answer this question, we calculates the spectra by assuming that the light trajectories are straight. Under such kind of calculations, the disk self-shadowing effect is nearly absent. Thus we conclude that the strong self-shadowing effect is due to the bending of light by the strong gravity. This is also sustained by the fact that for the same luminosity where the maximum number of $h/r$ is the same, the disk self-shadowing effect is more significant. This is illustrated in Figure [{\bf TBD}].

\subsection{Diversity in Spectral Evolution}

%%need a good graph here, to explain the diversity caused by shadowing
It was known that the spectral evolution of the X-ray binary deviates from the trend predicted in the SSD model
at both the low and high luminosities \citep{2007A&ARv..15....1D, 2004ApJ...601..428K}.  Although the deviation at the low luminosity states is often interpreted as the evaporation \citep{1994A&A...288..175M} of the optically thick disk into the optically thin ADAF \citep[Advection Dominated Accretion Flow,][]{ 1997ApJ...482..448N,1995ApJ...452..710N,1994ApJ...428L..13N},
the deviation at the high luminosity states is still mysterious.

The standard accretion disk model predicts the monotonic evolution of the disk peak energy with the luminosity,
i.e., the higher peak energy corresponds to the higher luminosity. This kind of spectral evolution is consistent with
the spectra evolution of X-ray binary at the medium luminosity states . However, beyond some critical luminosity $L_{\rm crit}$
%(where $l_{\rm crit}$ is in Eddington units $l=L_{\rm disk}/ L_{\rm edd}$), with the increase of disk luminosity,
the peak of the disk SED shifts to the low energy band \citep{2007A&ARv..15....1D}, with the increase of the disk luminosity.
This kind of spectral evolution is difficult to be understood in the SSD model.

To tackle this problem, various attempts have been made, which include a change of the color correction with the luminosity
\citep{2006ApJ...647..525D} as well as the effects of the optically thick advection.
One common difficulty with these explanations is that they all predicts a single critical luminosity $L_{\rm crit}$,
which is in contradiction with the diversity of the critical luminosity observed in the different sources \citep{2007A&ARv..15....1D}.

One natural explanation from our calculations is that the abnormal of the spectral evolution is due to
the self-shadowing of the accretion disks. The trend of the observed spectral evolution is consistent with our results
as shown in Figure \ref{spectral}, and the diversity of the critical luminosity $L_{\rm disk}/L_{\rm edd}$ in the different sources
might be due to the different inclination angle and black hole spin. According to our results,
the critical luminosity becomes smaller for the larger inclination angle or the black hole spin.

Our results also provide a diagnostic of the location of the shadowing medium.
If the shadowing happens at the inner region of the accretion disks, the critical luminosity is dependent on
black hole spin as well as the inclination angle. If the disk is shadowed by the matter distant from the black hole,
(e.g., equatorial wind; \citep{2008ApJ...683..389D}), the critical luminosity is dependent only on the black hole spin.

\subsection{The Spin of GRS 1915+105}
\begin{figure}[here!]
 \includegraphics[width=0.4 \textwidth]{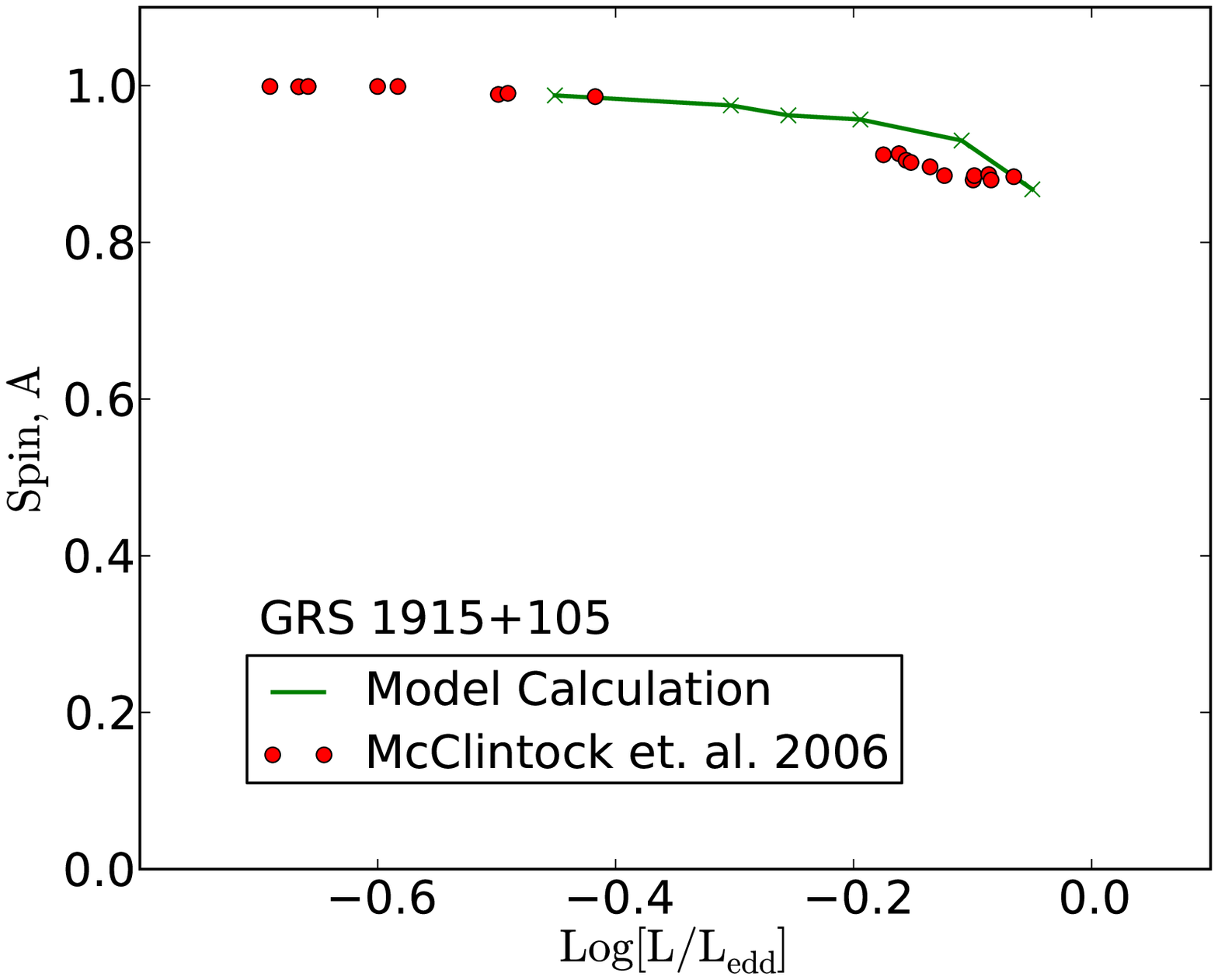}
  \caption{\label{1915_fit}
 Measured spin of GRS 1915+105 as a function of the disk luminosity.
The green line and dots are the results calculated in our model,
including the effects of the heat advection and the disk self-shadowing.
%using parameters appropriate for GRS 1915+105.
The red dots are the spin obtained in \citet{2006ApJ...652..518M} by
fitting the observed SED.
%Our model and observation agrees at both low
%and high luminosity. At medium luminosity, the disagreement may be caused by disk wind \citep{2008ApJ...683..389D}.
  }
 \end{figure}
The spin of GRS 1915+105 is debated in the literature. By fitting the disk blackbody component, \citet{2006ApJ...652..518M} found that the spin
is near extreme, while \citet{2006MNRAS.373.1004M} found the spin is moderate, with $a \sim 0.8$. One major difference between these two works
is that the spin obtained by \citet{2006ApJ...652..518M} is based on the spectra fitting for the relatively low luminosity ($L<0.3 L_{\rm edd}$) states,
while \citet{2006MNRAS.373.1004M} set no strict restriction on the luminosity states.  Interestingly, \citet{2006ApJ...652..518M} found that the measured
spin becomes smaller for the higher luminosity states.

To show how the spin can be recovered by the fitting procedure, we simulate the artificial data in our model which is added with componization model $\mathrm{comptt}$ \citep{1994ApJ...434..570T}, iron absorption edge, iron absorption line as well as galactic absorption, and then
we fit the data with the same procedure as used in \citet{2006ApJ...652..518M}. All the parameters are the same as those
in \citet{2006ApJ...652..518M}. The only difference in the artificial spectral fitting is that we replace the KERRBB2 model with KERRBB model, and set
the spectral hardening factor to be $1.7$.

Our results are shown in Figure \ref{1915_fit}. Without introducing the additional parameters, we are able to roughly reproduce the trend found
in \citet{2006ApJ...652..518M}. At both the low (around 1/4 of Eddington luminosity) and high (close to 1 Eddington luminosity) luminosity states,
our simulated result agrees well with the result of \citet{2006ApJ...652..518M}. When the disk luminosity is low enough, according to the
spectral model we have developed, the spin determination by \citet{2006ApJ...652..518M} is reliable. When the luminosity is high enough, the
disk shadowing makes the measured spin lower than the actual value. This luminosity dependence of the measured spin is shown up both in
our calculation and in the observational data. When the disk luminosity is medium (around $10^{-0.2}$ Eddington luminosity), our results
begin to diverge from the data. This may due to the photon scattering by the wind from the accretion disk as suggested by
\citet{2008ApJ...683..389D}. However, as our spectra fitting is based on the model of \citet{2006ApJ...652..518M}, so it is still
possible that different spectral model \citep[e.g. a low temperature comptonization component,][]{2006MNRAS.373.1004M} may lead to a different result.

%  \begin{figure}[here!]
%  \includegraphics[width=0.4 \textwidth]{1915.eps}
%   \caption{\label{1915}
%  Simulated spectra of GRS1915+105. The mass of black hole is $14 M_{\odot}$, the spin is $0.998$ and inclination angle is set to $66^{\circ}$. The spectral hardening factor $f_{\rm col}$ is set to 1.7. The accretion rate takes value $\dot M=\frac{1}{16} \dot M_{\rm edd}$, $\dot M=\frac{1}{4} \dot M_{\rm edd}$, $\dot M=\dot M_{\rm edd}$ respectively(correspond to $L\sim 0.32 L_{\rm edd}$,$1.28 L_{\rm edd}$, $5.12 L_{\rm edd}$). The symbol of lines is the same as in Figure \ref{spectral}. Parameters are the same as that suggested by \citet{2006ApJ...652..518M}.
%   }
%  \end{figure}
}
\section{Conclusions}
\label{recon}
In this work, we study the structure of the relativistic slim accretion
disks and their emergent spectra. Besides the effect of the photon trapping, we find that when the accretion rate is appropriate,
the energy trapped in the outer regions of the accretion disks can be re-radiated in the inner regions of the accretion disks.
Another main result of this work is that we check the validity of the relativistic SSD. We find that at low accretion rate,
such as $\dot M \sim 1/4$ for $a=0$ and $\dot M \sim 1/16$ for $a=0.98$, the relativistic SSD is still valid to describe the
disk structure, and the effect of heat advection will not significantly influence the global structure of the disk.
%
%Based on the global solutions of the disk, we model the emergent
%spectra from such accretion disks. We find that at the relatively low accretion rate such as $\dot
%M=1/4\dot M_{\rm edd}$, $a=0$ and $\dot M=1/16\dot M_{\rm edd}$, $a=0.98$(which correspond to $\sim 0.2$ of Eddington luminosity),
%both the effect of heat advection and the effect of disk self-shadowing are very small. In these cases, our
%results suggest that the SSD model is valid of describing the disk structure.
At relatively
high accretion rate such as $\dot M = 2 \dot M_{\rm edd}$, these effects will
significantly distort the emergent spectrum, and hence affect the measurement of the black hole spin. We also find that when the accretion
rate is high, the effect of the disk self-shadowing plays an even more important role than that
of the heat advection alone.

% We also checked in our numerical code that this kind of disk self-shadowed effect is a combination of thick geometry and light bending effect. By turning off the light bending effect in our calculation, the self-shadowing effect is highly suppressed. Thus our results suggest that the previous estimation of self-shadowing based on non-relativistic geometrical argument is inaccurate.

The emergent luminosity from the accretion disks is angular dependent, and the degree of anisotropy is dependent on the
accretion rate. In this work, we investigate the angular dependence of the observed luminosity. Due to the physical
similarities between the accretion disks of different
black hole mass, our result can be applied to estimate the radiation anisotropy of AGNs.
% and give tables and fitting formulas for direct comparison with the  observations. Our results can be tested by tracing the state evolution of the X-ray binaries \citep[e.g.][]{2004ApJ...601..428K,2004MNRAS.353..980K,2007A&ARv..15....1D}.

To check in principle the effect differences of spectra on estimation of spin, we create ``fake'' data sets with
the simulated data, and fit them with the KERRBB model using XSPEC. At
the relatively low accretion rate, our fitting results suggest that the
simple Keplerian model is reliable. At the relatively high accretion
rate, our fitting results suggest that self-shadowing effect will lead to significant under-estimation of
the black hole spin. {By making our calculation relevant to the case of GRS 1915+105, without introducing additional parameters, we are able to reproduce the decrease of measured spin with the increasing luminosity, which is found in \citet{2006ApJ...652..518M}. This finding suggests that disk
self-shadowing significantly shapes the spectra of GRS 1915+105.}

% Due to the complications in getting transonic disk solution, the spectral fittings based on the slim disks are seldom carried out. The transonic disk solution depends on the black hole mass, accretion rate and black hole spin. Thus to get disk solutions with reasonable parameter coverage, substantial amount of calculation is needed. However, the solutions of different black hole mass are related by some simple scaling relations at inner regions of the accretion disks. Thus a grid of data for parameter $a$ and $\dot M$ will give transonic disk solutions with reasonable parameter coverage. We are working on a table to achieve that, together with a friendly interface to give convenient access to that table.
{
There are several effects which we do not include in our calculations.
It was claimed that the disk may have no time to relax to the thermodynamic equilibrium \citep{2002ApJ...574..315O},
this effect will make our estimation of emergent flux (Equation (\ref{rad})) invalid and
make the disk thickness estimation inaccurate. In our calculations of the disk SED, we use a
simplified model of the local spectrum and neglect the effect of limb--darkening and returning irradiation.
Assuming a constant color correction may
make our calculated spectra inadequate to be compared to the observation \citep{2008ApJ...683..389D}, so we focus on
analyzing the physical effects we have introduced.
The limb-darkening effect will influence the spectra and the angular dependence of the emergent luminosity,
while the returning irradiation will have some effects on both the disk structure and the emergent spectra.
As \citet{2005ApJS..157..335L} have calculated, these two effects will contribute several percents to the final results.
Despite of these shortcomings, our results and the trend of the spectral evolution are still robust when the inclination angle is not too large.
}

\acknowledgments
We thank the anonymous referee for her/his constructive suggestions which
are very helpful for the improvement of this paper and thank Prof. Jian-min Wang and Mr. Yanrong Li for
discussions. G.-X. Li thanks Prof. Chris Done for discussion, and Mr. Teng Liu for help with the XSPEC software.
This work is partially supported by National Basic Research
Program of China (grant 2009CB824800), the National Natural Science
Foundation (grants10733010, 10673010, 10573016,
10773020, 10821302 and 10833002), the CAS (grant KJCX2-YWT03) and the Science and Technology Commission of Shanghai Municipality (10XD1405000).

\bibliography{bh1}
\end{document}